\documentclass[final,5p,times,twocolumn]{elsarticle}
\usepackage{amsfonts}
\usepackage{amsmath}
\usepackage{amsthm}
\usepackage{amssymb}
\usepackage{graphicx}
\usepackage{dcolumn}
\usepackage{bm}
\usepackage{bbding}
\usepackage{mathrsfs}
\usepackage{graphicx}
\biboptions{numbers,sort&compress}

\begin{document}

\begin{frontmatter}
\title{Uncertainty-induced quantum nonlocality}
\author{Shao-xiong Wu}
\author{Jun Zhang}
\author{Chang-shui Yu\corref{cor1}}
\ead{quaninformation@sina.com}
\cortext[cor1]{Corresponding author. Tel: +86 41184706201}
\author{He-shan Song}
\address{School of Physics and Optoelectronic Technology, Dalian University of
Technology, Dalian 116024, China}
\begin{abstract}
Based on the skew information, we present a quantity, uncertainty-induced quantum nonlocality (UIN) to measure the quantum correlation. It can be considered as the updated version of the original measurement-induced nonlocality (MIN) preserving the good computability but eliminating the non-contractivity problem.   For $2\times d$%
-dimensional state, it is shown that UIN can be given by  a closed form. In addition, we also investigate the maximal uncertainty-induced nonlocality.
\end{abstract}

\end{frontmatter}

\section{Introduction}

Quantum entanglement is the important resource for quantum protocols \cite
{rmp}. However, it is shown that some quantum tasks such as deterministic quantum
computation with one quantum bit (DQC1) \cite{DQC11,DQC12,DQC13}, quantum state merging \cite{statemerging1,statemerging2,statemerging3} and so on, can also display obvious
quantum advantages without any entanglement but possibly including quantum discord \cite {discord1,discord2,luo08}. This could be one of the important reasons why quantum discord has attracted so many attentions in the past years (see Ref. \cite{modi} and the references therein).

Quantum discord can be roughly understood as the disturbance on the state of interest induced by one-side local measurements \cite{discord1,discord2}. Of course, the different versions of quantum discord correspond to the different measures of the disturbance\cite{MIN06,MIN10,MID,1fanshu,deficit,modiprl,GD,Bures}. In the similar spirit, the measurement-induced nonlocality (MIN) is defined as the maximal change on a bipartite quantum state after a projective measurement which does not influence the local one-side density matrix \cite{MIN11}. MIN  has been studied in various fields \cite{ MINuse3,MINuse4,MINuse5,MINuse6, MINuse7,monogamy,MINHu,MINXi,MINGuo,MINbose,steering,remotestatecontrol,densecoding}, such as the dynamics \cite{ MINuse3,MINuse4,MINuse5,MINuse6, MINuse7}, the
monogamy \cite{monogamy} and so on.
However, like the geometric quantum discord \cite{GD}, the definition of MIN is also  based on the Hilbert-Schmidt norm, so it will inevitably suffer from the non-contractivity problem \cite{bujisuo,wenti1,wenti2,wenti3}.  That is, MIN  could be increased by the non-unitary evolution on the subsystem without measurements and
by adding an extra product state. For geometric quantum discord, even though lots of new definitions have
been proposed \cite{1fanshu,1fanshu2,Bures,remedy,rescale,lqu} to avoid the non-contractivity problems, only the local quantum uncertainty
(LQU) \cite{lqu} based on the skew information \cite{skew1,skew2} overcomes the non-contractivity without losing the computability. For MIN, can we also use skew information to solve the problems with the computability preserving?

In the Letter, we give the positive answer. We find that if  the considered local observable on subsystem $A$ commutes with the reduced density matrix $\rho_a$, we can obtain a MIN-like nonlocality measure, \textit{i.e.}, uncertainty-induced quantum nonlocality (UIN). It is found that UIN will vanish for the states without MIN, so UIN
can be considered as a revised version of MIN with contractivity. In particular,  for $2\times d$-dimensional
states, UIN  can also be given by a closed form. In addition, we also investigate the maximal
uncertainty-induced nonlocality. The Letter is organized as follows. In Section 2, we give the definition of
UIN and present the
closed form for $2\times d$ dimensional systems. In Section 3, we investigate
the maximal
uncertainty-induced nonlocality. At the end, the
summary is given.
\section{Uncertainty-induced quantum nonlocality}
\subsection{The definition}
To begin with, we  would like to give a brief introduction about local quantum uncertainty \cite{lqu}. The skew information of a density $\rho$ and a local observable $K^{\Lambda }$ is defined by
\begin{eqnarray}
\mathcal{I}(\rho ,K^{\Lambda })=-\frac{1}{2}\mathrm{Tr}[\sqrt{\rho },K^{\Lambda }]^{2},\label{eq1}
\end{eqnarray}
where $K^{\Lambda }=K_{a}^{\Lambda }\otimes \mathbb{I}_{b}$  with $K_{a}^{\Lambda }$ is Hermitian operator on subsystem $A$ with
non-degenerate spectrum $\Lambda $. For a $2\times d$-dimensional state $\rho$, LQU is defined by
\begin{eqnarray}
\mathcal{U}_{A}(\rho ) &=&\min_{K^{\Lambda }}\mathcal{I}(\rho ,K^{\Lambda })\notag \\
&=&1-\lambda _{\max }(W_{ab}),  \label{lqu}
\end{eqnarray}%
where $\lambda _{\max }$ is the maximal eigenvalue of $3\times 3$ symmetric matrix $W_{ab}$ with
the entries $(W_{ab})_{ij}=\mathrm{Tr}\{\sqrt{\rho}(\sigma _{i}\otimes \mathbb{I}%
_{b})\sqrt{\rho}(\sigma _{j}\otimes \mathbb{I}_{b})\}$, $i,j=x,y,z$.

In order to obtain a MIN-like nonlocality, we have to require that the local commuting observable $K^{C}=K_{a}^{C}\otimes \mathbb{I}_{b}$ commutes with the reduced density matrix $\rho _{a}$, where $K_{a}^{C}$ is Hermitian operator on $A$ with
nondegenerate spectrum $\Lambda'$. Thus the UIN can be  defined by the maximal skew
information of the state $\rho $ and local commuting observable $K^{C}$, which is given in a rigorous way as follows.

\textbf{Definition 1.} \textit{Uncertainty-induced quantum nonlocality for a bipartite state $\rho$ is given by}
\begin{eqnarray}
\mathcal{U}_{C}(\rho )=\max_{K_{C}}\mathcal{I}(\rho ,K^{C}).
\end{eqnarray}

Next, we would like to list the good properties of UIN, by which one can find that UIN can be considered as a updated version of MIN.

(1). $\mathcal{U}_{C}(\rho )$ is invariant under local unitary operation. This can be seen as follows. Let local unitary operators $U_{A}\otimes U_{B}$ operate on the state $\rho$, then the UIN becomes
\begin{eqnarray}
&&\mathcal{U}_{C}((U_{A}\otimes U_{B})\rho (U_{A}\otimes U_{B})^{\dagger })
\notag \\
&=&\max_{K_{a}^{C }}\mathcal{I}((U_{A}\otimes U_{B})\rho (U_{A}\otimes
U_{B})^{\dagger },K_{a}^{C}\otimes \mathbb{I}_{b})  \notag \\
&=&\max_{K_{a}^{C }}\mathcal{I}(\rho ,(U_{A}\otimes U_{B})^{\dagger
}(K_{a}^{C}\otimes \mathbb{I}_{b})(U_{A}\otimes U_{B})  \notag \\
&=&\max_{\widetilde{K_{a}^{C}}}\mathcal{I}(\rho ,\widetilde{K_{a}^{C}}%
\otimes \mathbb{I}_{b})  \notag \\
&=&\mathcal{U}_{C}(\rho ),
\end{eqnarray}%
where $\widetilde{K_{a}^{C}}=U_{A}^{\dagger }K_{a}^{C}U_{A}$ commutes
with $\rho _{a}$ because $[K_a^{C},U_A\rho U_A^{\dagger}]=0$.

(2). $\mathcal{U}_{C}(\rho )$ is nonincreasing (contractive) under local
operation on subsystem $B$. As we know, the skew information $\mathcal{I}(\rho ,K)$ is
contractive under local operations on $B$. That is, if the local operation is denoted by $\varepsilon(\cdot)$, one has $\mathcal{I}%
(\varepsilon (\rho ),K)\leqslant \mathcal{I}(\rho ,K)$. Assume $K^{0}$ is the optimal observable for $\varepsilon
(\rho )$ such that the corresponding Eq. (3) holds, then one can arrive at
\begin{eqnarray}
\mathcal{U}_{C}(\varepsilon (\rho )) &=&\mathcal{I}(\varepsilon (\rho
),K^{0})  \notag\\&\leqslant& \mathcal{I}(\rho ,K^{0})  \notag\\&\leqslant& \mathcal{U}_{C}(\rho ).
\end{eqnarray}%

(3). For product state $\rho_p =\rho _{a}\otimes \rho _{b}$, $\mathcal{U}%
_{C}(\rho_p )=0$. This is obvious, because the operator $K^{0}=K_{a}^{0}\otimes \mathbb{I}_{b}$ commutes with  $\rho _{a}\otimes \rho _{b}$, if  $K_{a}^{0}$ commutes with $\rho _{a}$.

\subsection{The closed form for $2\otimes d$-dimensional system}
Next we will give the closed form of UIN for an arbitrary $2\otimes d$-dimensional quantum state.

\textbf{Theorem 1.} \textit{For any bipartite $2\otimes d$-dimensional state $\rho$, UIN is given by
\begin{eqnarray}
\left\{
\begin{array}{cc}
\mathcal{U}_{C}(\rho )=1-\lambda _{\min }(W_{ab}), & \mathbf{r}=0 \\
\mathcal{U}_{C}(\rho )=1-\frac{1}{\left\vert \mathbf{r}\right\vert ^{2}}%
\mathbf{r}W_{ab}\mathbf{r}^{T}, & \mathbf{r}\neq 0%
\end{array}%
\right. , \label{dingli1}
\end{eqnarray}%
where $\left\vert \cdot \right\vert $ denotes vector norm, $W_{ab}$ is
defined in Eq.(\ref{lqu}).}

\textbf{Proof}. For any bipartite $2\otimes d$-dimensional state, the reduced
density matrix of subsystem $A$ is denoted by $\rho _{a}=\frac{1}{2}(\mathbb{I+}%
\mathbf{r\cdot \sigma }_{a})$ with $\mathbf{r}$ the Bloch vector. Analogous to Ref. \cite{lqu},  we directly adopt the considered
local  traceless observable $K_{a}^{C}=%
\mathbf{n\cdot \sigma }_{a}$  with $\vert\mathbf{n}\vert=1$. Thus $[K_{a}^{C},\rho _{a}]=0$ will directly lead to \begin{eqnarray}
\mathbf{r}\times \mathbf{n}=0.  \label{duiyi}
\end{eqnarray}

Eq.(\ref{duiyi}) can be divided into two cases:

1). $\mathbf{r}=0$, which implies that the reduced density matrix is identity
matrix, \textit{i.e.}, $\rho _{a}=\frac{1}{2}\mathbb{I}$. In this case, any operator
commutes with the reduced density matrix $\rho _{a}$. So $\mathbf{n}$ can be
an arbitrary normalized vector, so the UIN can be given by
\begin{eqnarray}
\mathcal{U}_{C}(\rho ) &=&\max (1-\mathbf{n}W_{ab}\mathbf{n}^{T})  \notag \\
&=&1-\lambda _{\min }(W_{ab}),  \label{lmqu}
\end{eqnarray}%
where $\lambda _{\min }(W_{ab})$ is the minimal eigenvalue of $W_{ab}$, and $%
\mathbf{n}$ is the corresponding eigenvector.

2). $\mathbf{r}\neq 0$, this will lead to $\mathbf{n}=c\mathbf{r}$, $c\in
\mathbb{R}$. One can obtain that%
\begin{eqnarray}
\mathcal{U}_{C}(\rho ) &=&1-\mathbf{n}W_{ab}\mathbf{n}^{T}  \notag \\
&=&1-\frac{1}{\left\vert \mathbf{r}\right\vert ^{2}}\mathbf{r}W_{ab}\mathbf{r%
}^{T}.  \label{disanzhongqingkuang}
\end{eqnarray}
The second $'=' $ is attributed to $\left\vert \mathbf{n}\right\vert ^{2}=1$, \textit{i.e.},  $\left\vert
c\right\vert ^{2}=\frac{1}{\left\vert \mathbf{r}\right\vert ^{2}}$. Eq. (\ref{lmqu}) and Eq. (\ref{disanzhongqingkuang}) give the closed form of $\mathcal{U}_{C}(\rho )$. $\Box $

\textbf{Corollary 1}. \textit{For a $2\otimes d$-dimension pure state $\left\vert \psi
\right\rangle $,\ UIN reduces to entanglement monotone
\begin{eqnarray}
\mathcal{U}_{C}(\left\vert \psi \right\rangle )=2(1-\mathrm{Tr}\rho
_{a}^{2}).  \label{chuntai}
\end{eqnarray}
}

\textbf{Proof}. It is easy to find that $W_{ab}$ has only one eigenvalue when $%
\rho $ is a pure state. The eigenvector of $W_{ab}$ is just
the local Bloch vector $\mathbf{r}$,
\begin{eqnarray}
\mathbf{v} &=&(\left\langle \psi \right\vert (\sigma _{1}\otimes \mathbb{I}%
_{b})\left\vert \psi \right\rangle ,\left\langle \psi \right\vert (\sigma
_{2}\otimes \mathbb{I}_{b})\left\vert \psi \right\rangle ,\left\langle \psi
\right\vert (\sigma _{3}\otimes \mathbb{I}_{b})\left\vert \psi \right\rangle
)  \notag \\
&=&(r_{1},r_{2},r_{3})=\mathbf{r}.
\end{eqnarray}%
The eigenvalue of $W_{ab}$ is the squared norm of $\mathbf{v}$, \textit{i.e.},
\begin{eqnarray}
\lambda (W_{ab})=|\mathbf{v}|^{2}=r_{1}^{2}+r_{2}^{2}+r_{3}^{2}=2\mathrm{Tr}%
\rho _{a}^{2}-1.\label{eq12}
\end{eqnarray}
When $\mathbf{r}=0$, substitute Eq.(\ref{eq12})
into Eq.(\ref{lmqu}), Eq.(\ref{chuntai})
will be obtained immediately. When $\mathbf{r}\neq 0$, it is easy to find that
\begin{eqnarray}
\mathcal{U}_{C}(\left\vert \psi \right\rangle ) &=&1-\frac{1}{\left\vert
\mathbf{r}\right\vert ^{2}}\mathbf{r}W_{ab}\mathbf{r}^{T}  \notag \\
&=&1-\frac{1}{\left\vert \mathbf{r}\right\vert ^{2}}\mathbf{rv}^{T}\mathbf{vr%
}^{T}  \notag \\
&=&1-\left\vert \mathbf{r}\right\vert ^{2}  \notag \\
&=&2(1-\mathrm{Tr}\rho _{a}^{2}).  \label{chuntai2}
\end{eqnarray}%
In Eq.(\ref{chuntai2}), we've used the fact that $W_{ab}=\mathbf{v}^{T}\mathbf{v}
$, $\mathbf{v=r}$ and $|\mathbf{v}|^{2}=2\mathrm{Tr}\rho _{a}^{2}-1$. $%
\Box $

\textbf{Corollary 2}. \textit{UIN can be explained as the maximal squared Hellinger
distance between $\rho $ and $K^{C}\rho K^{C}$.}

\textbf{Proof}. In \cite{lqu}, the authors pointed out that LQU can be
reinterpreted geometrically as the minimal squared Hellinger distance
between $\rho $ and $K^{\Lambda }\rho K^{\Lambda }$. Here, replacing $%
K^{\Lambda }$ by $K^{C}$, we will obtain that
\begin{eqnarray}
\mathcal{I}(\rho ,K^{C}) &=&1-\mathrm{Tr}\{\sqrt{\rho }K^{C}\sqrt{\rho }%
K^{C}\}  \notag \\
&=&1-\mathrm{Tr}\{\sqrt{\rho }\sqrt{K^{C}\rho K^{C}}\}  \notag \\
&=&D_{H}^{2}(\rho ,K^{C}\rho K^{C})
\end{eqnarray}%
where $D_{H}^{2}(\rho ,K)=\frac{1}{2}\mathrm{Tr}\{(\sqrt{\rho }-\sqrt{K}%
)^{2}\}$ is the squared Hellinger distance, $K^{C}=K_{a}^{C}\otimes \mathbb{I}_{b}$
, and $K_{a}^{C}=\mathbf{n\cdot \sigma }_{a}$ with $|%
\mathbf{n}|^{2}=1$ and $[K_{a}^{C},\rho _{a}]=0$.
So maximize over the local commuting observable $K^{C}=K_{a}^{C}\otimes
\mathbb{I}_{b}$, we will have a geometric interpretation of the UIN using
Hellinger distance. $\Box $

\subsection{Example}

According to the preceding results, one can find that UIN can be considered
as an updated version of MIN without non-contractivity problem. Here, we give
a concrete example to show this advantage of UIN.

Let's consider a state
 \begin{eqnarray}
\rho_{ab} =\left[
\begin{array}{cccc}
0.4205 & 0.0805 & 0.3278 & 0.0966 \\
0.0805 & 0.1757 & 0.0564 & 0.0840 \\
0.3278 & 0.0564 & 0.2808 & 0.0615 \\
0.0966 & 0.0840 & 0.0615 & 0.1230%
\end{array}%
\right] ,
\end{eqnarray} which is randomly generated by Matlab 6.5.
Suppose the subsystem $B$ undergoes an
amplitude damping channel with the final state given by $\varepsilon (\rho_{AB}
)=\sum_{k=0}^{1}(\mathbb{I}\otimes E_{k})\rho_{ab} (\mathbb{I}\otimes
E_{k}^{\dagger })$ where $E_{0}=\left\vert 0\right\rangle \left\langle
0\right\vert +\sqrt{1-\gamma }\left\vert 1\right\rangle \left\langle
1\right\vert ,E_{1}=\sqrt{\gamma }\left\vert 0\right\rangle \left\langle
1\right\vert $. The MIN and the UIN for $\varepsilon$ are
shown in Fig. 1. Obviously, MIN suffers from the non-contractivity problem,
while UIN is contractive.

\begin{figure}[th]
\centering
\includegraphics[width=0.8\columnwidth]{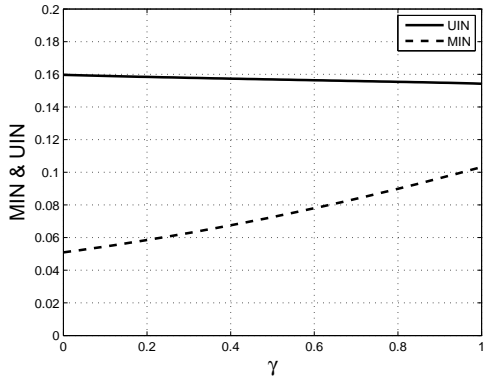}
\caption{(Dimensionless) The UIN and the MIN for the state $\varepsilon(\rho_{ab})$ \textit{vs.} the rate $\gamma$.  The solid line stands for the
UIN, and the dashed line corresponds to the MIN.}
\end{figure}

\section{Maximal uncertainty-induced nonlocality}

MIN is an important measure of nonlocality. However, MIN is not the dual definition of geometric quantum discord without any restriction on the measurements \cite{zhangjun} .   For integrity, we would like to present the direct dual definition of LQU.  For the UIN, the local commuting
observable $K^{C}=K_{a}^{C}\otimes \mathbb{I}_{b}$ is restricted to satisfy
 $[K_{a}^{C},\rho_a]=0$, so the direct dual measure to the LQU should be defined as the maximum skew information between the state and any possible local observable. In other words, $K_{a}^{C}$ should be optimized over the set of all local observables. To do so, we  can define the maximal uncertainty-induced nonlocality (MUIN) as follows. \begin{eqnarray}
\mathcal{U}_{M}(\rho ) &=&\max_{K^{\Lambda }}\mathcal{I}(\rho ,K^{\Lambda })  \notag \\
&=&1-\lambda _{\min }(W_{ab}),
\end{eqnarray}%
where $K^{\Lambda }$ is  the local observable on subsystem $A$ defined in Eq. (\ref{eq1}).
One can easily check that MUIN also has some good properties, such as the invariance under local unitary
operations, the contractivity under local operation on subsystem $B$, reducing to
entanglement monotone when $\rho $ is pure state. In addition, $%
\mathcal{U}_{M}(\rho )$ is equal to UIN in the case of $\mathbf{r}=0$, while $%
\mathcal{U}_{M}(\rho )$ is bigger than UIN in other cases.
However, MUIN may not be a good measure for the nonlocality, for example, it does not vanish for the product states.

\section{Conclusion}

We have introduced a measure UIN based on skew information. UIN is the
maximal skew information between state and the local commuting observable
$K^{C}$ and can be reinterpreted as the maximal squared Hellinger distance
between $\rho $ and $K^{C}\rho K^{C}$.  UIN can be considered
as an updated version of MIN without the non-contractivity problem. It has a good computability, because
for a $2\times d$-dimension state, UIN can be given by a closed form. In addition, we also studied the maximal uncertainty-induced nonlocality.

In \cite{MIN11}, the authors pointed out that the non-contractivity problem of MIN can be remedied in terms of Helinger distance or Bures distance. The Bures distance, which is
monotonic,  Riemannian and  can be connected with Uhlmann fidelity, seems to be
a good candidate \cite{pra2013}. However it may be difficult to obtain analytical solution
for a general state (much like the geometric discord based on trace distance \cite{1fanshu,1fanshu2}
and Bures distance \cite{Bures}). Anyway, it still deserves our  further endeavor.

\section*{Acknowledgement}

This work was supported by the National Natural Science Foundation of China,
under Grants No.11375036 and 11175033 and by the Fundamental Research Funds of the
Central Universities, under Grant No. DUT12LK42.

\end{document}